\documentclass[preprint, pra, amsmath, aps,showpacs]{revtex4}
\usepackage{color}
\newcommand{\bea}{\begin{eqnarray} }
\newcommand{\eea}{\end{eqnarray}}
\newcommand{\bean}{\begin{eqnarray*}}
\newcommand{\eean}{\end{eqnarray*}}
\newcommand{\nn}{\nonumber \\}
\def\bfg#1{{\mbox{\boldmath $#1$}}}
\def\A{{\bf A}}
\def\B{{\bf B}}
\def\b{{\bf b}}

\def\E{{\bf E}}

\def\g {{\bf g}}

\def\R {\bf R}

\def\v{{\bf v}}

\def\x {{\bf x}}
\def\X {{\bf X}}

\def\btimes{~{\bf \times}~}
\def\bnabla{{\bf \nabla}}
\def\bcdot{~{\bf \cdot}~}
\newcommand{\lbs}{\left (}
\newcommand{\rbs}{\right )}
\newcommand{\lbm}{\left\lbrack}
\newcommand{\rbm}{\right\rbrack}

\newcommand{\bp}{\indent$\bullet\hspace*{5mm}$}
\def\od#1,#2{\frac{d#1}{d#2}}
\def\odz#1,#2{\frac{d^2#1}{d{#2}^2}}
\def\pd#1,#2{\frac{\partial #1}{\partial #2}}
\def\pdz#1,#2{\frac{\partial^2 #1}{\partial {#2}^2}}

\def\pdd#1,#2{\frac{\partial^3 #1}{\partial {#2}^3}}
\def\pdv#1,#2{\frac{\partial^4 #1}{\partial {#2}^4}}
\def\pdzz#1,#2,#3{\frac{\partial^2 #1}{\partial {#2}\partial{#3}}}

\def\eq#1{Eq.~(\ref{#1})}
\def\eqn#1{(\ref{#1})}
\usepackage{graphicx}
\usepackage{dcolumn}
\usepackage{bm}

\begin{document}


\preprint{Published in Phys. Plasmas {\bf 30} 104702 (2023)}
\bibliographystyle{unsrt}
%
%
\title{
Response to “Comment on ‘Modification of Lie's transform perturbation theory for 
charged particle motion in a magnetic field’” [Phys. Plasmas 30, 104701 (2023)] 
}
%
%
\author{Linjin  Zheng\email{lzheng@austin.utexas.edu}}
\affiliation{Institute for Fusion Studies,
University of Texas at Austin,
Austin, TX 78712}
\date{\today}

\begin{abstract}

 Dr. Brizard's comment on my work  is based on a conceived procedure
 that does not come from my work.
 The defense of  his claim that
 the modification of the so-called standard Lie's transform theory
 is unnecessary is also unsupported. 
 This response reveals in detail the inconsistency issues  in the so-called standard Lie's transform theory
 by  analyzing both its results and root causes. 
  The problem in the so-called standard Lie's transform 
theory is beyond the issue to take into account
  the ordering difference  between the guiding center motion and gyromotion. 
The inconsistent commutation of derivative and
limit causes another issue. 
 Besides, the so-called standard Lie's transform formulation
leads to an unnecessarily lengthy and tedious
derivation process for  a one or two page task under the singular (or renormalized) formalism described in my paper.

\end{abstract}


\maketitle

Introduction of Lie's transform perturbation theory to plasma physics is an important contribution. 
Dr. R. G. Littlejohn and Dr. J. R. Cary led a foundation-laying effort in this field
and Dr. Brizard et al. also made further important developments. 
However, the system with a fast varying coordinate, the gyrophase in the plasma physics application,
belongs to a singular perturbation problem. The so-called standard Lie's transform theory
for example in Refs. \cite{littlejohn83} and  \cite{brizard95}
met the ordering inconsistency issue, although the missing lower order effect is somehow  
recovered in the higher order equation. 
Therefore, in my paper a singular (or renormalized)
Lie's transform perturbation theory is developed. 
The development from the regular perturbation theory to the singular one is a
 pattern in many fields of physics and classical mechanics.
 It is a necessary and important progress. It affects not only the theory of a charged 
 particle motion, but also Lie's algebra in general.
 Nonetheless, it is merely  a drop among numerous pioneer contributions both in mathematics and 
 physics, in our field specially by  Dr. R. G. Littlejohn, Dr. J. R. Cary, including
 Dr. Brizard et al.  
 In view of this, I have avoided using ``correction" in the title of my paper,
but using ``modification".  In fact, the commutation of derivative and limit  in the
derivation of Lagrangian one form
is indeed incorrect mathematically. As pointed out in my paper,  Dr. Cary and Dr. Brizard actually
made an important contribution in developing the renormalized perturbation theory in the
direct approach,\cite{cary09} although Dr. Brizard seems not realize it (see the later discussion).

As pointed out in my paper,\cite{zheng} the so-called standard
 Lie's transform theory for charged particle motion was first reported in Ref. \cite{littlejohn83}.
However, the details are omitted, leaving them upon private request.  Dr. Brizard later 
 detailed and extended  it in the Appendix B of Ref. \cite{brizard95}, which is important 
 for introducing this approach to  fusion community   and also helpful for the further development
 for example in my paper.\cite{zheng}
Since the details are needed in the discussion, Dr. Brizard's work in Ref. \cite{brizard95}
is used in the following response.  The response contains two parts: the counter arguments against Dr. Brizard's defense on the so-called standard Lie's transform theory
and the defense of my own work. Both the results and root causes are analyzed.

\section{The results show the ordering inconsistency issues in the so-called standard Lie's transform theory}

\label{rev}

In this section,  the ordering inconsistency issues in the so-called standard Lie's transform theory 
in Refs. \cite{littlejohn83} and  \cite{brizard95} are pointed out by analyzing its resulting Lagrangian
formula. Comparison between my work and Dr. Brizard's one is made to show 
how the issues are solved in my approach.

Let me first outline the results in my paper.\cite{zheng}
 Dr. Brizard started his comment with the claim that
``Zheng never defines an ordering parameter in his critique of standard Lie-transform perturbation theory and his proposed modification of Lie-transform perturbation theory is completely unnecessary."
This apparently does not reflect the fact. 
The orderings have been clearly specified in my paper in the paragraph after Eq. (2): 
``In these analyses, the gyrofrequency
$\Omega=eB/mc$ is assumed to be high, i.e., $(v/R) /\Omega \sim (\partial/\partial t) /\Omega
\sim {\cal O} (\epsilon) \ll 1$, where
$R$ is the scale of electromagnetic field, which is  larger than the Larmor radius by an order of magnitude. 
It is also assumed that $v_\E =|\E\times\B|/B^2 \ll v$."

The zeroth and first order Lagrangian one forms in my work \cite{zheng} are given as follows
 \bea
 {^d}\Gamma^{(0)} &=&     \frac{e}{\epsilon mc} \A  \bcdot d\X ,
 \label{ze0}
\\
 {^d}\Gamma^{(1)}  &=&  u\b \bcdot d\X +\frac{mc}e\mu d \zeta - \lbs  \frac {u^2}2+\mu B +
 \frac{e}m \varphi\rbs  dt .
\label{ze1}
\eea
As pointed out in  the line describing Eq. (50),
 ``combining the contributions from $^d\Gamma^{(0)}$  and  $^d\Gamma^{(1)}$, one obtains".
 \bea
 {^d}\Gamma &=& \epsilon\lbm \lbs    \frac{e}{\epsilon mc} \A +u\b\rbs \bcdot d\X +\frac{mc}e\mu d \zeta - \lbs  \frac {u^2}2+\mu B +
 \frac{e}m \varphi\rbs  dt + {\cal O}(\epsilon) \rbm .
\label{r1}
\eea
In particular, the term $\frac{mc}e\mu d \zeta$ is recovered in the first oder equation, \eq{ze1}. 

Next, the results in Brizard's work in Ref. \cite{brizard95} are detailed.
This is necessary since Dr. Brizard has used improper and obscure interpretation of
the ordering tags, such as  $e/m\to \epsilon e/m$, or $m\to \epsilon m$.
Detailing the order-by-order results in Dr. Brizard's paper can  unambiguously reveal the ordering scheme
used in his work, i.e., the so-called standard Lie's transform theory in
Refs. \cite{littlejohn83} and  \cite{brizard95}.

For simplicity,  the  rotation is
assumed to be small. 
In the second line
in subsection 1 ``Zeroth order perturbation analysis" in Appendix B, Dr. Brizard obtains
\bea
\Gamma_0 = \A(\X)\bcdot d\X.
\label{gamma0}
\eea
In subsection 2 ``First order perturbation analysis" in Appendix B, he obtains Eq, (B18) as follows
\bea
\Gamma_1 &=& u\b\bcdot d\X- H_1  dt.
\label{gamma1}
\eea
In subsection 3 ``Second order perturbation analysis" in Appendix B, he obtains Eq, (B30) as follows
\bea
\Gamma_2 &=&\mu \tilde \R \bcdot d\X   +\mu d \zeta  - H_2 dt,
\label{gamma2}
\eea
where $\tilde \R\sim -( 1/ 2)\b\bcdot\bnabla \btimes \b \b $.
Note that the constants $e/m$ and $c$ are taken out in Eqs. \eqn{gamma0} - \eqn{gamma2}.
Combining these results Dr. Brizard finally obtains Eq.(B39) as follows
 \bea
 \Gamma &=&\epsilon\lbm   \lbs    \frac{e}{\epsilon mc} \A +u\b\rbs \bcdot d\X+  \epsilon 
  \frac{mc}e\mu \tilde \R \bcdot d\X +\epsilon
 \frac{mc}e\mu d \zeta - \lbs  \frac {u^2}2+\mu B +
 \frac{e}m \varphi\rbs  dt\rbm
\label{rbriz0}
\eea
with the constants $e/m$ and $c$ being added back.

Note that there is a key difference between my result in \eq{r1} and Dr. Brizard's one in \eq{rbriz0}.
In my work, the term $\frac{mc}e\mu d \zeta$ appears in the first order equation, \eq{ze1},
and is of the same order as $u\b \bcdot d\X$, i.e.,
\bea
\frac{mc}e\mu d \zeta\sim u\b \bcdot d\X.
\label{ordz}
\eea
Instead, in Dr. Brizard's work,  the term $\frac{mc}e\mu d \zeta$ appears in the second order 
equation, \eq{gamma2}, and is considered to be of the same order as 
$ \frac{mc}e\mu \tilde \R \bcdot d\X$, i.e.,
\bea
\frac{mc}e\mu d \zeta\sim  \frac{mc}e\mu \tilde \R \bcdot d\X.
\label{ordb}
\eea

Which result is right can be seen by trivial ordering analyses.
My result in \eq{ordz} can be confirmed as follows
\bea
\frac{ (mc/e)\mu \dot \zeta}{u\b\bcdot \dot\X}&\sim& \frac{ (mc/e)(v_\bot^2/B) (d \zeta/dt)}{u\b\bcdot (d\X/dt)}
\nn
&\sim& \frac{(\epsilon v_\bot^2/(eB/m  c)) (d \zeta/dt)}{u\b\bcdot (d\X/dt)}\sim\frac {v_\bot^2}{  u^2} \sim 1;
\label{r3}
\eea    
while Dr. Brizard's result in \eq{ordb} can be disapproved as follows 
\bea
\frac { \epsilon(mc/e)\mu \tilde \R \bcdot d \X} { \epsilon(mc/e)\mu d \zeta}&\sim&
 \frac{ \dot \X /|X|}{ \dot \zeta}\sim\frac{transit/bounce~frequency }{gyrofrequency} \ll 1.
\label{esti}
\eea
These prove trivially that the ordering scheme in Dr. Brizard's work is inconsistent, while my work is right. 
The ordering estimate in \eq{esti} indicates that Dr. Brizard actually treats
the gyrofrequency as the same order as the transit/bounce frequency of the guiding
center motion.

Furthermore, using my results in  \eq{r1}, one can recover the well-known the guiding center drift velocity\cite{littlejohn83,zhengb}
\bea
\dot \X &=& u\b +\epsilon \lbs \frac {u^2\b\btimes(\b\bcdot\bnabla\b)}{\Omega} 
+\frac{\mu B \b\btimes\bnabla \ln B} { \Omega}   \rbs;
\label{zvd}
\eea
while using Brizard's results in \eq{rbriz0}, the guiding center drift velocity becomes
\bea
\dot \X &=& u\b +\epsilon \lbs \frac {u^2\b\btimes(\b\bcdot\bnabla\b)}{\Omega} 
+\epsilon\frac{\mu B \b\btimes\bnabla \ln B} { \Omega}   \rbs.
\label{bvd}
\eea
With $\epsilon$ present in the last term in \eq{bvd},  Dr. Brizard's result wrongly 
concludes  that the grad-$B$ drift is negligible as compared to
the curvature drift.

\section{The root causes for the ordering inconsistency in the so-called standard Lie Transform theory}

In this section, let me discuss 
the root causes for the ordering inconsistency in the so-called standard Lie transform theory
in Refs. \cite{littlejohn83} and  \cite{brizard95}.
First, the ordering consideration is discussed in order to explain the necessity
of the  singular perturbation theory.\cite{zheng}
Before detailing the comparison between the regular and singular perturbation theories,
an apparent inconsistent treatment in the so-called standard Lie's transform theory is pointed out
to convince the readers why the modification of  so-called standard theory is necessary. 
 In the end, the modification of the regular perturbation theory by the singular perturbation theory in my paper\cite{zheng}
 is detaled.

\subsection{Ordering consideration in the singular perturbation theory in my work}

\label{so}

In this subsection, we review the ordering consideration in the singular perturbation theory developed 
in my work.\cite{zheng}
First, as pointed in my paper, one should note the ordering difference between the temporal variation of
gyrophase and that of  other phase space coordinates:
\bea
d\zeta = dZ^-/\epsilon,
\label{zo1}
\eea
where the supperscript (or subscript) ``$-$" has been introduced to denote the phase space coordinates with
$d\zeta$ excluded.
Equation \eqn{zo1} just indicates that the gyrofrequency is much larger than the transit/bounce frequency of guiding center motion.
Furthermore, note that the guiding center transformation till the first order is given as follows
\bea
z^\mu = Z^\mu+\epsilon g_1^\mu,  ~~\hbox{with}~~\g_1^\X = -\bfg{\rho}.
\eea
Although $g_1^\mu$ is of order $\epsilon$, the ordering for $dz^\mu$ is different:
\bea
dZ^- \sim |d\g^\X_1|  \sim |d\bfg{\rho}|\sim \left|\pd {\bfg{\rho}},\zeta d\zeta\right|.
\label{zo2}
\eea
This is just an example for the first order case. Higher order cases need also to be carefully treated. 

Furthermore, without commuting the derivative and limit,  one obtains the modified transform rule
for the Lagrangian one form as detailed in my paper\cite{zheng}
\bea
(L_g\bar \Gamma)_\mu &=&  g^\nu\lbs \partial_\nu\gamma_\mu    -\partial_\mu\gamma_\nu \rbs
\nn&&
-  g^\nu\lbm \underbrace{\lbs \partial_\nu\gamma_\delta\rbs\lbs \partial_\mu g_1^\delta\rbs }_{new~1}  
   -\underbrace{\lbs\partial_\mu\gamma_\delta \rbs \lbs \partial_\nu g_1^\delta\rbs }_{new~2}   \rbm+\cdots.
\label{zt}
\eea
Because of the ordering in \eq{zo2}, one can see that the correction in the second term can be of the same
order as the first term and therefore cannot be neglected. In particular, the term ``new~1", 
$ g^\nu \lbs \partial_\nu\gamma_\delta\rbs\lbs \partial_\mu g_1^\delta\rbs$,  simply 
gives rise to the term $\bfg{\rho} \bcdot \bnabla \A \bcdot  \pd \bfg{\rho},\zeta d\zeta$.
Noting the ordering in \eq{zo2}, this term is not of order $\rho/R\sim\epsilon$, but of order unity $\epsilon^0$,
as compared with the first term (also proportional to $g^\nu\sim\rho)$ on the right hand side of \eq{zt}, i.e., the transform formula for one form
in the so-called standard Lie's transform theory in Refs.
\cite{littlejohn82} and \cite{cary83}.

\subsection{An apparent inconsistent treatment in the so-called standard Lie's transform theory}

\label{ex}

Before detailing the comparison between the regular and singular perturbation theories,
 let me first point out an apparent inconsistent treatment in the so-called standard 
Lie's transform  theory in   Refs. \cite{littlejohn83} and  \cite{brizard95}
to convince the readers why the modification of  so-called standard theory is necessary. 

In my work,\cite{zheng} the Lagrangian one form $\Gamma_\mu$ is distinguished from
$^d\Gamma_\mu =  \Gamma_\mu dZ^\mu$, where no summation over $\mu$ is assumed.
This distinction is not always made in the so-called standard Lie's transform theory.
For example, in Ref. \cite{cary83} Dr. Cary define the Lagrangian one form $\Gamma_\mu$
without $d Z^\mu$ included, while  Dr. Brizard in Ref.  \cite{brizard95} with $d Z^\mu$ included.
They are related  in the so-called standard 
Lie's transform  theory  in Refs. \cite{littlejohn83} and  \cite{brizard95} as follows:
\bea
{^{Brizard}\Gamma^{(i)}} =~ {^{Cary}\bar \Gamma^{(i)}} \bcdot d\vec Z,
\label{brca}
\eea
where the superscrit $(i)$ denotes the order.
Mixing the notations between $\Gamma_\mu$ and $^d\Gamma_\mu$ is ok without considering the ordering. However, when the ordering in \eq{zo1} is taken into account,
one can find that \eq{brca}   should be modified to
\bea
{^{Brizard}\bar \Gamma^{(i)}} =~ {^{Cary}\Gamma_{\zeta}^{  (i+1)}}  d\zeta +~{^{Cary}\bar \Gamma_{-}^{(i)}} \bcdot d\vec Z^-. 
\label{brcam}
\eea
This is related to the correction pointed in my paper\cite{zheng} 
that the ordering difference between the temporal variation of
gyrophase and that of the other phase space coordinates needs to be taken into account.
This example is sufficient to disapprove Dr. Brizard's claim that
 the modification of the so-called standard Lie's transform theory
 is unnecessary.

\subsection{Modification of the regular perturbation theory by the singular perturbation theory in my paper}

Using the ordering consideration for the singular perturbation approach reviewed in Sec. \ref{so}, 
we now explain in detail how the regular perturbation theory   in Refs. \cite{littlejohn83} and  \cite{brizard95}
is modified order by order by the  singular perturbation theory developed in my paper.\cite{zheng} 

Dr. Brizard's zeroth and first order starting equations are\cite{brizard95}
\bea
^d\Gamma^{(0)} &=& dS^{(0)}  +  \frac{e}{mc} A_\mu(Z)  dZ^\mu,
\label{cb0}
\\
^d\Gamma^{(1)} &=& dS^{(1)}  -L^{conv}_1 ~{^d}\gamma^{(0)}  + {^d}\gamma^{(1)}.
\label{cb1}
\eea
We then describe how they are modified for ordering consistency in my work.\cite{zheng}

We first examine the modification to the lowest order equation, \eq{cb0}. Noting the ordering in \eq{zo2}, one can find
that \eq{cb0} needs to be modified to Eq. (42) in my paper as follows\cite{zheng} 
\bea
^d\Gamma^{(0)} &=& dS^{(0)}  +  \frac{e}{mc} A_\mu(Z)  dZ^\mu +   \underbrace{\frac{e}{mc}A_\mu(Z)  dg_1^\mu}_{new},
\label{zb0}
\eea
Here, a new term is added as compared to \eq{cb0}
(the sign of this new term is corrected). The new term is introduced because $dg_1^\mu$ can be of
order unity due to the presence of $d\zeta$ component. 
Nevertheless, the contribution  of this new term is negligible. Equation \eqn{zb0} yields \eq{ze0} in Sec. \ref{rev},
i.e., Eq.(43) in my paper.\cite{zheng}

We next examine the modification to the first order equation, \eq{cb1}. 
Noting Eqs. \eqn{zo1}-\eqn{zt} in subsection \ref{so},
one can find
that \eq{cb1} needs to be modified to Eq. (44) in my paper as follows\cite{zheng} 
\bea
^d\Gamma^{(1)} &=& dS^{(1)}  -L^{conv}_1 ~{^d}\gamma^{(0)}  + {^d}\gamma^{(1)} 
+  \underbrace{g_1^\mu d v_\mu}_{new~1}  +  \underbrace{v_\mu dg_1^\mu}_{new~2}
\underbrace{- g_1^\lambda  \pd \gamma^{(0)}_{\mu},{Z^\lambda} dg_1^\mu}_{new~3}
\nn
&=& dS^{(1)}  -L^{conv}_1 ~{^d}\gamma^{(0)}  + {^d}\gamma^{(1)} 
\underbrace{- g_1^\lambda  \pd \gamma^{(0)}_{\mu},{Z^\lambda} dg_1^\mu}_{new~3}.
\label{zb1}
\eea
Here, three new terms are added as compared to \eq{cb1}. The first new term is derived from
the contribution of $ -L^{conv}_1 ~{^d}\gamma^{(1)}$, i.e., the contribution $ g_1^\X \pd \v_\bot,\zeta d\zeta$. 
The second new term is derived from $^d\gamma^{(1)}$, in which $dg_1^\mu$ is kept noting
the ordering in  \eq{zo2}. The third new term is derived from the term ``new ~1" in \eq{zt}.
The combined effect from terms ``new~1" and ``new~2" is negligible since the dominant 
effect from $d\zeta$ component forms a total derivative. Therefore, only term ``new~3" remains.
Here, I would like to point out that there is a correction in Eq.(44) in my paper. The term $-v_\mu dg_1^\mu$
should be left out as Eq. (3.152) in my book\cite{zhengb} (or replaced by the terms  ``new~1" and ``new~2"). 
This is just a correction in the intermediate step due to typing overlook and does not affect the final results.
Equation \eqn{zb1} yields \eq{ze1} in Sec. \ref{rev}.  
Combining the zeroth and first order
contributions, Eq.(50) in my paper is obtained.\cite{zheng}

Let us discuss this a little bit further. The Lie's transform perturbation theory in Refs.  \cite{littlejohn82} and \cite{cary83}
uses the phase space description, which extends  the conventional $ 2N$ canonical coordinates $(q_n; p_n)$ to
the $ 4N$ phase space coordinates $(q_n, p_n; \dot q_n,\dot p_n)$. The phase space description is very interesting, but
it brings in extra variables and equations. The terms ``new~1" and ``new~2" are actually the products of the
extension to phase-space coordinates and therefore cancel each other eventually. It is the 
term ``new~3" that represents the finite Larmor radius effect from changing to the guiding center description,
i.e., $\A(\x)  = \A(\X+\bfg{\rho}) = \A(\X) + \bfg{\rho}\bcdot\bnabla \A \bcdot \bfg{\rho}$, 
which gives rise to the grad-B drift in \eq{zvd}.   This can also be seen clearly from the direct formulation in Ref. \cite{cary09}. 
Therefore, strictly speaking, the formulation in Refs.  \cite{littlejohn83} and  \cite{brizard95}
has not caught the actual source term for the grad-B drift. It is the consequence 
of inconsistent commutation between the derivative and limit.

These show how the regular Lie's transform theory, or the so-called standard Lie's transform theory,  in Refs. \cite{littlejohn83} and  \cite{brizard95}
is modified by the  singular perturbation theory developed in my paper
by taking into consideration the ordering scheme in Sec. \ref{so}.\cite{zheng}

\section{Dr. Brizard's comment   is based on a conceived procedure
 that does not come from my work.}

 Dr. Brizard's comment on my work  is based on his conceived procedure
 that does not come from my work.
Here are point by point responses:

\noindent \bp  {\sf Dr. Brizard's comment:}

``In his critique of standard Lie-transform perturbation
analysis, and without explicitly displaying the dimension-less ordering parameter $\epsilon$ upon which it is to be based,
Zheng [1] mistakenly proceeds to compare the guiding-center Lagrangians (4) and (5), derived with different
renormalization orderings, and concludes that, when the
guiding-center Lagrangian (5) is truncated at first order, the term 
$\epsilon^2\frac{mc}e\mu d \zeta$ disappears, while the term
 $\epsilon\frac{mc}e\mu d \zeta$ remains in the guiding-center Lagrangian (4),
although it is still a second-order term compared to the
lowest order term appearing at $\epsilon^{-1}$."

\noindent {\sf --- Response:} 
First, I point out that
 orderings are only meaningful when comparing terms with the same dimensions. Based
on this basic principle,  one can see that
Eq.(4) and Eq.(5) in Dr. Brizard's  comment are actually the  same things. 
Labeling $L'_{gc}$ as order $\epsilon$ is unsound. Taking the extreme,  for example,
comparing to infinity,  $L'_{gc}$ is zero; while comparing to zero,  $L'_{gc}$ is infinity.
Ordering only has a relative meaning.
What is more, as pointed out in my paper they are identically wrong
because they are based on the incorrect one-form transformation rule.
Subtracting two identical and identically wrong equations  to produce something 
meaningful has not been ``proceeded" in my paper.

\noindent \bp {\sf Dr. Brizard's comment:}

``However, Zheng
seems to be unaware that the guiding-center Lagrangian
(4), which was derived without Lie-transform perturbation method by Cary and Brizard [6] in what Zheng
calls the direct method, was also derived by Lie-transform
perturbation method by Littlejohn [4], Brizard [5], and
Tronko and Brizard [13]."

\noindent  {\sf --- Response:} This just indicates that Dr. Brizard does not take the credit I mentioned in my paper
that Dr. Cary and Dr. Brizard made an important contribution in developing the renormalized perturbation theory in the
direct approach. The direct approach in Ref. \cite{cary09} obtains  \eq{r1} as reviewed in Sec. 3.2 in my book,\cite{zhengb}    while Dr. Littlejohn and Dr. Brizard's Lie's transform formulation obtains \eq{rbriz0}  as  claimed in Dr. Brizard's comment.

\noindent \bp {\sf Dr. Brizard's comment:}

After Eq. (8) in Dr. Brizard's comment, he argues that the combination
\bea
\frac{mc}e\mu \lbs \delta \dot \zeta -\epsilon   \R \bcdot \dot \X\rbs
\eea
in the second order equation, \eq{gamma2}, has to have $\delta =\epsilon$  for the gyrogauge invariance.

\noindent  {\sf --- Response:} As shown  in \eq{esti}, there is  ordering  inconsistent if $\delta =\epsilon$. 
As pointed out in Sec. \ref{ex}, the term containing $\dot\zeta$ should be raised to the first order equation, \eq{gamma1}.
The argument of ordering tags or the terminology of so-called gyrogauge invariance does not justify
the apparent ordering inconsistency.

\section{Conclusions}

In conclusion,  Dr. Brizard's comment on my work  is based on a conceived procedure
 that does not come from my work.
 The defense of  his claim that
 the modification of the so-called standard Lie's transform theory
 is unnecessary is also unsupported. 
 The problem in the so-called standard Lie's transform 
theory is beyond the issue to take into account
  the ordering difference  between the guiding center motion and gyromotion. 
The inconsistent commutation of derivative and
limit causes another issue. 
 Besides, the so-called standard Lie's transform formulation
leads to an unnecessarily lengthy and tedious
derivation process for  an one or two page task under the renormalized formalism described in my paper.

Note that if the transform rule for Lagragian one form  has to be corrected,
the existing derivation for gyrokinetic equation by the so-called standard Lie's transform 
theory\cite{bri07,cb} certainly needs to be corrected. Also, for the theory of charged particle motion
the modification is not only the first order equation, \eq{gamma1}, but also the second order one, \eq{gamma2}.
 
The regular perturbation method somehow does pick up the missing lower effects through
higher order analyses. However, it is not good mathematically in ordering consideration. 
Therefore, as in other fields of physics and classical mechanics, as soon as the renormalized 
perturbation formalism is
proposed, the regular approach for singular problem should be gradually 
given up.

This research is supported by Department of Energy Grants DE-FG02-04ER54742.

\end{document}